# Vacuum gauge from ultrathin MoS$_2$ transistor

A. Di Bartolomeo*[1,2], A. Pelella[1,2], A. Grillo[1,2], F. Urban[1,2], L. Iemmo[1,2], E. Faella[1,2], N. Martucciello[2], and F. Giubileo[2]

[1]Dipartimento di Fisica "E.R. Caianiello", via Giovanni Paolo II, 132, 84084, Fisciano, Salerno, Italy
[2]CNR-Spin, via Giovanni Paolo II, 132, 84084, Fisciano, Salerno, Italy

*Email: adibartolomeo@unisa.it

**Abstract.** We fabricate monolayer MoS$_2$ field effect transistors and study their electric characteristics from $10^{-6}$ Torr to atmospheric air pressure. We show that the threshold voltage of the transistor increases with the growing pressure. Hence, we propose the device as an air pressure sensor, showing that it is particularly suitable as a low power consumption vacuum gauge. The device functions on pressure-dependent O$_2$, N$_2$ and H$_2$O molecule adsorption that affect the n-doping of the MoS$_2$ channel.

## 1. Introduction

Following the great success of graphene [1–4], several families of atomically thin materials have emerged in the past decade and have been dominating the material research scenario ever since [5, 6]. In particular, two-dimensional (2D) transition metal dichalcogenides (TMDs) have attracted a lot of attention due to several promising properties for electronic, optoelectronic, energy, catalysis and sensing applications [7–9]. TMDs consist of a "sandwich" structure (layer) with a transition-metal sheet located in between two chalcogen sheets and possess unique properties such as energy bandgap tunable by the number of layers (form 0 to about 2.2 eV), good mobility up to few hundreds cm$^2$V$^{-1}$s$^{-1}$ photoluminescence, broadband light adsorption, surface without out-of-plane dangling bonds that allows the fabrication of hetero-structures, high strength with Young's modulus up to 300 GPa, exceptional flexibility, and thermal stability in air [10–12]. They are produced by mechanical or liquid exfoliation, chemical vapor deposition (CVD), molecular beam epitaxy, pulsed laser deposition, etc. [13].

Molybdenum disulfide (MoS$_2$), formed by covalently bonded S–Mo–S sequences in easy-to-exfoliate 2D layers, held together by weak van der Waals forces, is the most investigated TMD [14, 15]. It is a semiconductor with 1.2 eV indirect bandgap in the bulk form that widens up to 1.8-1.9 eV and becomes direct in the monolayer. It is a promising material for field-effect transistors (FETs) with high performance and on/off ratio [16–18], sensitive broadband photodetectors [19, 20], catalysis [21], chemical and biological [22–25] or strain and pressure sensors [23, 26].

Microscopic pressure sensors that can rapidly detect small pressure variations are of high demand in robotic technologies, human–machine interfaces, electronic skin, sound wave detection, and health monitoring devices. Pressure sensors are very important in many other fields, such as automobiles, aircrafts, well drilling, and medical applications.

The exceptional mechanical properties of $MoS_2$ nanosheets [27, 27] have inspired their application as ultrathin diaphragms capable of large deflection deformations at low pressure to achieve high sensitivity in pressure sensors. For instance, a thin and sensitive diaphragm is attached onto one end face of a cleaved optical fiber to form an extrinsic Fabry-Perot interferometric structure that detects the applied pressure through the measurement of the deflection deformation of the diaphragm. Fabry–Perot ultrasensitive pressure sensors with nearly synchronous pressure–deflection responses have been fabricated using few-layer $MoS_2$ films. Compared to conventional diaphragm materials (e.g., silica, silver films), they have allowed to achieve three orders of magnitude higher sensitivity (89.3 nm $Pa^{-1}$) [28].

Highly sensitive pressure sensors have been fabricated by integrating a conductive microstructured air-gap gate with $MoS_2$ transistors. The air-gap gate is used as the pressure-sensitive gate for 2D $MoS_2$ transistors to reach pressure sensitivity amplification to $\sim 10^3$–$10^7$ $kPa^{-1}$ at an optimized pressure regime of $\sim 1.5$ kPa [29].

Due to the atomic thickness, the electrical properties of two-dimensional materials are highly affected by ambient gases and their pressure variations. The adsorbed gas modifies the electron states within 2D materials changing their electrical conductivity. Owing to the low adsorption energy the process can be reversible.

Specifically, it has been demonstrated that $MoS_2$ conductivity can be enhanced or suppressed by gases such as $O_2$, $CH_4$, $NO_2$, NO, $NH_3$, $H_2S$, etc. [22, 30, 31]. Therefore, few- and single-layer $MoS_2$ nanosheets have been investigated for gas and pressure sensing in devices with fast response speed, low power consumption, low minimum pressure detection limits and excellent stability. For instance, few-layer $MoS_2$ back-gate field effect transistors, fabricated on $SiO_2$/Si substrate with Au electrodes, have been demonstrated as resistor-based $O_2$ sensors with sensing performance controllable by the back-gate voltage. Remarkably, these devices have been applied to determine $O_2$ partial pressure with a detectability as low as $6.7 \times 10^{-7}$ millibars at a constant vacuum pressure and proposed as a vacuum gauge [32].

In this paper, we fabricate $MoS_2$ back-gate field effect transistors using $MoS_2$ nanosheets grown by chemical vapor deposition (CVD) on $SiO_2$/Si substrate and measure their electrical characteristics at different air pressures. We show that the threshold voltage of the transistor increases with the increasing pressure. We ascribe such a feature to pressure-dependent adsorption of electronegative oxygen, nitrogen and water molecules, which decrease the n-doping of the $MoS_2$ channel and hence increase the threshold voltage of the transistor. We propose to exploit the dependence of the transistor current on the air pressure to realize vacuum gauges with wide dynamic range and low power consumption.

## 2. Experimental

The MoS$_2$ monolayer flakes were grown by CVD on a heavily doped Si substrate covered by 285 nm SiO$_2$, spin coated with 1% sodium cholate solution. The molybdenum needed for the growth was provided by a saturated ammonium heptamolybdate (AHM) solution, which was annealed at 300 °C under ambient conditions to turn AHM into MoO$_3$. The substrate and the AHM solution were placed in a three-zone tube furnace, along with 50 mg of S powder, positioned upstream in a separate heating zone. The zones containing the S and the AHM were heated to 150 °C and 750 °C, respectively. After 15 min of growth, the process was stopped, and the sample was cooled rapidly.

MoS$_2$ nanoflakes with different shapes and thicknesses, depending on both the local stoichiometry and temperature, were formed [33]. An example is shown in Figure 1a).

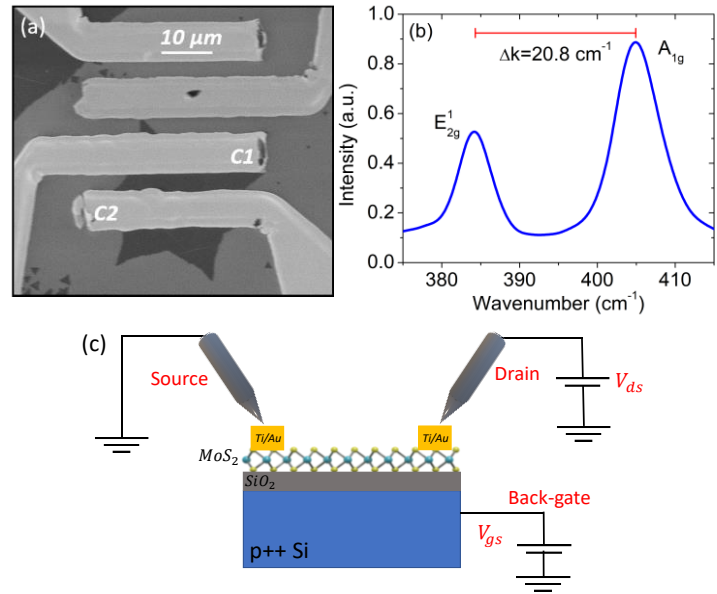

**Fig. 1**. **a)** SEM image showing the star-like MoS$_2$ nanoflake used as the channel of the back-gate transistor with Ti/Au contacts. The channel width and length are 28.0 μm and 4.4 μm, respectively. **b)** Raman spectrum of the MoS$_2$ nanoflake with $E_{2g}^1$-$A_g^1$ wavenumber separation corresponding to a monolayer. **c)** MoS$_2$ FET schematic with biasing circuits used for the electrical characterization.

We used optical microscope inspection, with contrast calibrated to approximately estimate the number of layers, to identify MoS$_2$ nanoflakes suitable for the transistor fabrication. A standard e-beam lithography and lift-off process was applied to evaporate Ti (10 nm) and Au (40 nm) bilayers on the flake for the formation of the source and drain electrodes. The back-gate electrode was formed by scratching the Si substrate surface and dropping silver paste.

The SEM top view of a typical device, fabricated using a star-like nanoflake, is shown in Figure 1a). Figure 1b) shows the Raman spectrum of the flake under 532 nm laser excitation. The wavenumber difference, $\Delta k \approx 20.8$, between the $E^1_{2g}$ (in-plane optical vibration of S atoms in the basal plane) and $A^1_g$ (out-of-plane optical vibration of S atoms along the c axis) indicate a monolayer [34, 35].

Figure 1c) displays the schematic cross-section of the device and the circuits used for the electrical characterization of the transistor in common source configuration. The electrical measurements were carried out inside a cryogenic probe station with fine pressure control (Janis ST 500), connected to a Keithley 4200 SCS (source measurement units, Tektronix Inc.), at room temperature.

## 3. Results and discussion

Figures 2a) and 2b) show the output, $I_{ds} - V_{ds}$, and the transfer $I_{ds} - V_{gs}$ characteristics of the MoS$_2$ transistor measured in high vacuum and at room temperature. As often observed in MoS$_2$ and other 2D-material based devices, the output characteristic exhibits an asymmetric behavior for positive and negative drain biases. As we have demonstrated elsewhere, such a feature is caused by the different contact area as well as by a difference in the Schottky barrier height at the two contacts resulting from local MoS$_2$ processing or intrinsic defects [36, 37]. The transfer characteristic shows a normally-on, n-type transistor. The intrinsic n-type conduction is typical of MoS$_2$ and is mainly due to S vacancies [38]. Compared to similar devices reported in the literature, the transistor shows good metrics in terms of on/off ratio of $10^8$ at $\pm 60$ V, on-current $\sim 0.3 \frac{\mu A}{\mu m}$, subthreshold swing of 3.5 $\frac{V}{decade}$ and $\mu = \frac{L}{W C_{ox} V_{ds}} \frac{dI_{ds}}{dV_{gs}} \approx 1.2$ cm$^2$V$^{-1}$s$^{-1}$ (L and W are the channel length and width, $V_{ds}$ is the source-drain bias and $C_{ox} = 12.1$ nFcm$^{-2}$ is the SiO$_2$ capacitance per area) [39–41].

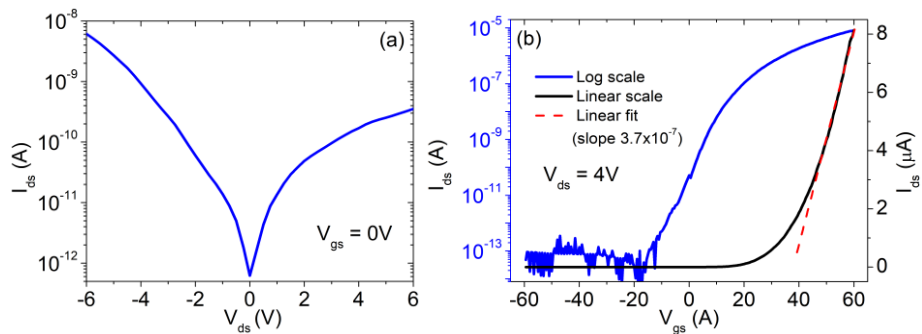

**Fig. 2**. **a)** Output and **b)** transfer (on linear and logarithmic scale) characteristics of the device between C1 and C2 contacts measured a room temperature and $10^{-6}$ Torr pressure. The dashed red line is a linear fit used to evaluate the channel field effect mobility.

The result of transfer characteristic measurements at different pressures, P, from high vacuum to atmospheric pressure and back to $10^{-6}$ Torr, is displayed in Figure 3a). The increasing air pressure causes a left-shift of the transfer curve and therefore an increase of transistor threshold voltage, $V_{th}$. The threshold voltage is here defined as the x-axis intercept of the straight lines fitting the $I_{ds} - V_{ds}$ curves in the current range 1-100 nA. The effect is reversible, in fact the device returns to the pristine state when the high vacuum is restored, as shown by the dash-dot grey line in Figure 3a). We note that the effect of air pressure on the channel conductance, which could result in the dramatic transformation of n-type to p-type conduction when passing from high vacuum to atmospheric pressure, has been reported also for other 2D TMDs materials such as $WSe_2$ or $PdSe_2$ [11, 42]. The effect is usually reversible although it has been found that an aging can occur in specific TMDs, such as $PdSe_2$, after a long (>20 days) air exposure at atmospheric pressure[43].

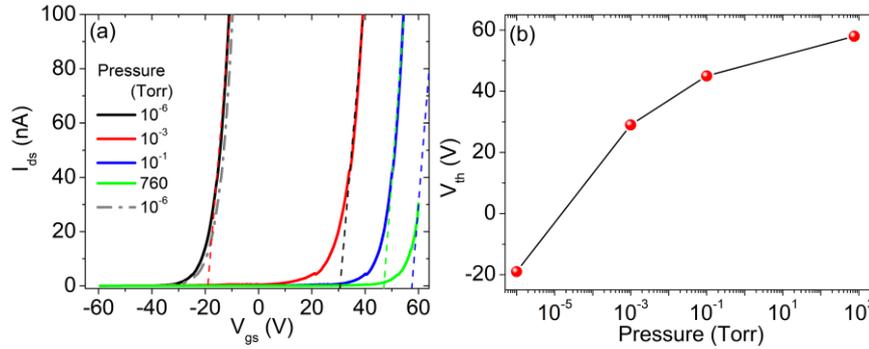

**Fig. 3**. **a)** Transfer characteristics on linear scale for increasing pressure from high vacuum to atmospheric. The dashed lines are the fitting straight lines used to evaluate the transistor threshold voltage. The dash-dot gray line is obtained after that the high vacuum is restored. **b)** Threshold voltage as a function of the pressure.

The monotonic $V_{th} - P$ behaviour, shown in figure 3b), suggests that the transistor can be used as pressure sensor, with maximum sensitivity up to $\frac{dV}{d(\log_{10} P)} \approx 13 \frac{V}{decade}$ at lower pressures, where the $V_{th} - P$ curve is steeper. Besides the higher sensitivity, the duty cycle of the device increases when operated in vacuum because of the suppressed air aging effect. Therefore, the sensor is best suited as a vacuum gauge. Moreover, low current of 1 nA or less is needed to monitor the $V_{th}$ variation, which implies that of the sensor can be operated in low power-consumption regime.

To investigate the working principle of the device, we measured the transfer characteristics over a gate voltage loop (backward and forward sweep) in air and vacuum. Figure 4a) shows the appearance of a hysteresis that reduces with the decreasing pressure. Hysteresis is a well-known phenomenon in transistors with 2D material channels and has been attributed to charge trapping in intrinsic defects of the 2D material, in the gate dielectric and in adsorbate molecules [38, 40, 44–46]. The reducing hysteresis with pressure confirm that adsorbates play an important role in the device under study.

Owing to their high electronegativity, molecular $O_2$, $N_2$ and $H_2O$, adsorbed on $MoS_2$ surface (Figure 4b)), can withdraw electrons from the channel causing the observed increase of the threshold voltage, i.e. of the gate voltage needed to enable conduction in the transistor channel. Absorption occurs particularly at S vacancy sites and the absorption/desorption rate obviously depends on the air pressure the $MoS_2$ nanosheet is exposed to, thus enabling its monitoring.

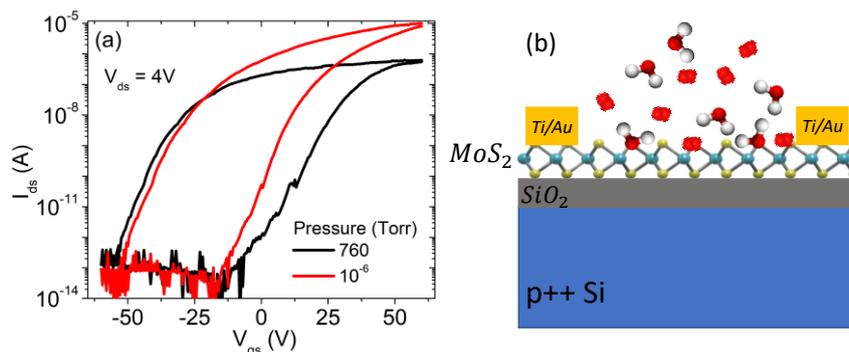

**Fig. 4**. **a)** Transfer characteristics showing a hysteresis between the forward and reverse $V_{gs}$ sweeps. The hysteresis width reduces with the decreasing pressure from atmospheric to $10^{-6}$ Torr. **b)** Schematic showing the adsorption of molecular $O_2$, $N_2$ and $H_2O$ which, being electronegative, cause the decrease of the electron density in the transistor channel and an increase of the threshold voltage.

## 3. Conclusion

We have fabricated and electrically characterized monolayer $MoS_2$ field effect transistors. We have found that the threshold voltage of the transistor increases monotonously with the air pressure, as effect of reducing n-doping caused by adsorption of electronegative $O_2$, $N_2$ and water. Therefore, we have proposed the transistor as an air pressure sensor, highlighting its suitability as a vacuum gauge with long duty cycle and low power consumption.